\documentstyle[12pt]{article}
\title{Black-hole information puzzle: A generic string-inspired approach}
\author{Hrvoje Nikoli\'c \\
Theoretical Physics Division, Rudjer Bo\v{s}kovi\'{c} Institute, \\
P.O.B. 180, HR-10002 Zagreb, Croatia \\
{\normalsize hrvoje@thphys.irb.hr} \\
\makebox[1in]{} \\
}
\date{\today}
\begin{document}
\maketitle
\begin{abstract}
Given the insight steming from string theory,
the origin of the black-hole (BH) information puzzle is traced back to the
assumption that it is physically meaningful to trace out the density
matrix over negative-frequency Hawking particles.
Instead, treating them as virtual particles necessarily absorbed by the
BH in a manner consistent with the laws of BH thermodynamics, and tracing
out the density matrix only over physical BH states, 
the complete evaporation becomes compatible with unitarity.
%As the physical massless BH state is unique,
%the final state of positive-energy outgoing Hawking particles is a pure
%state with a thermal-like distribution of particle numbers and energies.  
\end{abstract}
\vspace*{0.5cm}
%Keywords: Black-hole information; String theory \newline
%\vspace*{0.9cm}
%\pacs{04.70.Dy}
%\\

\section{Introduction}

The semiclassical description of
black-hole (BH) radiation \cite{hawk1} suggests that an
initial pure state evolves into a final mixed thermal state \cite{hawk2}. 
A transition prom a pure to a mixed state is incompatible with
unitarity of quantum mechanics (QM), which constitutes the famous
BH information puzzle. The attempts to restore unitarity can be divided
into two types (for reviews, see, e.g., \cite{har}).
%\cite{har,pag,gid,str}).
In the first type, the black hole does not evaporate
completely, but ends in a Planck-sized remnant that contains the information
missing in the Hawking radiation. The problem is that such a light object
should contain a huge amount of information, which seems unphysical.
In particular, light objects that may exist in a huge number of different
states should have a huge probability for creation in various physical
processes, which, however, is not seen in experiments.
A variant of the remnant scenario is the creation of a baby-universe not observable from our universe, but such an idea remains rather speculative.
In the second type, the black hole evaporates completely, but the radiation is not exactly thermal. Instead, there
are some additional subtle correlations among radiated particles. 
It is argued that this requires a sort of nonlocality not present
in standard quantum field theory (QFT), suggesting that quantum gravity
should contain some new nonlocal features. 

The most promising candidate for a consistent theory
of quantum gravity is string theory. Indeed,
it provides new insights on BH thermodynamics 
(see, e.g., \cite{hor,peet} for reviews). In particular, it provides 
a unitary description of BH radiation and offers a
microscopic explanation of the BH entropy proportional to the surface.
It also contains some nonlocal features that might explain
the desired deviation from exact thermality.
Nevertheless, the theoretical description of the mechanism 
of BH radiation in string theory
(see, e.g., \cite{das}) 
seems completely different from that in the conventional semiclassical
theory, so it remains difficult to see where exactly
the semiclassical analysis fails. Thus, it would be desirable to understand 
a generic property that a large class of models of quantum gravity,
including string theory, should
possess in order to save the unitarity of BH radiation. 
The aim of this paper is to find such a generic resolution of the BH information puzzle, without using any explicit model of quantum gravity.
We find that neither a new sort of nonlocality (for the case of complete
evaporation) nor a huge amount of information in a light remnant
(for the case of a remnant scenario) is needed. In fact, we find that
no new unexpected property of physical laws is required. Instead, the standard
rules of QM applied to black holes
in a generic and intuitively appealing manner turn out to be sufficient.

\section{Physical insights}

\subsection{Pure thermal states and decoherence}

First, let us observe that a thermal distribution of particles is not
necessarily incompatible with a possibility that these particles are
in a pure state. For a simple example, consider a single 
quantum harmonic oscillator
(with the frequency $\omega$) in the state
%\begin{equation}\label{e0}
$ |\psi\rangle = \sum_n f_{\omega,n}|n\rangle $,
%\end{equation}
where $n=0,1,2,\ldots$ and
\begin{equation}\label{e2.1}
f_{\omega,n}=\sqrt{1-e^{-\beta\omega}} e^{-\beta\omega n/2} .
\end{equation}
Clearly, $|\psi\rangle$ is a pure state. Yet, the probabilities of
different energies $E=\omega n$ are proportional to $e^{-\beta E}$,
which corresponds to a thermal distribution with the temperature
$T=1/\beta$.
The density matrix $\rho=|\psi\rangle \langle\psi|$ can be written as
\begin{equation}\label{rho}
\rho=\sum_n |f_{\omega,n}|^2 |n\rangle \langle n| +
\sum_{n\neq n'} f_{\omega,n}f_{\omega,n'} |n\rangle \langle n'| .
\end{equation}
The first (diagonal) term represents the usual mixed thermal state. The second
(off-diagonal) term is responsible for the additional correlations steming from the fact that the state is pure. When a simple system (in this case, the single harmonic oscillator) interacts with an environment with a large number of 
unobserved degrees of freedom, then, in practice, the presence of the second term is unobservable. Thus, for all practical purposes, 
the state can be described by the first term only. In QM this is known as the
phenomenon of decoherence (for a review see, e.g., \cite{schlos}).
Thus, decoherence provides a mechanism for an effective transition
from a pure to a mixed state
\begin{equation}\label{decoh}
|\psi\rangle \langle\psi| \stackrel{\rm decoher}{\longrightarrow}
\sum_n |f_{\omega,n}|^2 |n\rangle \langle n| .
\end{equation}
It does not involve any violation of unitarity at the fundamental
level.

\subsection{The role of negative-frequency particles in semiclassical
and fully quantum black holes}

As we shall see, the observations above will play a role in our resolution
of the BH information paradox. Indeed,
the role of decoherence in BH thermodynamics has already been discussed in
\cite{kief}. Nevertheless, decoherence is not the main part of our resolution.
To see the true origin of the BH information puzzle, we start from the fact
that standard semiclassical analysis based on the Bogoliubov transformation 
describes Hawking radiation as particle creation in which the initial
vacuum $|0\rangle$ transforms to a squeezed state \cite{bd}
\begin{equation}\label{e1}
|0\rangle \stackrel{\rm squeeze}{\longrightarrow}
|\psi\rangle_{\rm squeeze} ,
\end{equation}
where
\begin{equation}\label{e2}
|\psi\rangle_{\rm squeeze}=\prod_{\omega}\sum_n f_{\omega,n}(M)
|n_{-\omega}\rangle \otimes |n_{\omega}\rangle ,
\end{equation}
and, for massless uncharged spin-0 particles, $f_{\omega,n}(M)$ are given by (\ref{e2.1}) with $\beta\equiv 8\pi M$, where $M$ is the BH mass.
The product is taken over all possible positive values of $\omega$.
The state $|n_{\omega}\rangle$ represents on outgoing state containing
$n_{\omega}$ particles, each having frequency $\omega$, so that their
total energy is $E=\omega n_{\omega}$. Similarly, $|n_{-\omega}\rangle$
represents $n_{-\omega}$ ingoing particles, each having {\em negative} frequency
$-\omega$.
In our notation, the direct product $\otimes$ separates the 
inside states on the left from the outside states on the right.
At this level the total energy is not yet conserved, as 
the energy of the negative-frequency states is also positive,
in the sense that the sign of their energy is the same as that of
the interior matter determining the BH mass $M$.
The conservation of energy is provided by another mechanism, namely 
by renormalization of the energy-momentum tensor implying a
flux of negative energy across the horizon into the
black hole \cite{bd}. The overall effect is that the BH mass 
decreases, such that the total energy is conserved.
However, owing to the creation of negative-frequency particles that carry
information, the information content of the black hole {\em increases}
despite the fact that its mass decreases. Does it contradict the 
first law of BH thermodynamics? Not necessarily, if the BH entropy
proportional to the BH surface (and thus to $M^2$) is interpreted
merely as the part of BH information that is available to the outside 
observer. However, string theory suggests a very different interpretation
of BH entropy -- the entropy associated with counting of 
the internal degrees of freedom of the black hole, independent on the
knowledge of an outside observer. Thus, from the string-theory point 
of view, the information carried by the negative-frequency particles should be {\em unphysical}. Indeed, the physical mechanism of BH radiation in string
theory does not rest on the Bogoliubov transformation, and hence
does not lead to creation of particles in the BH interior \cite{das}.
Thus, our idea is to modify the semiclassical description of particle creation,
in a manner that removes the negative-frequency particles from physical
states.

For states $|n_{-\omega}\rangle$ we find convenient to
introduce a negative effective ``renormalized"
energy $E=-\omega n_{-\omega}$, without changing the information content
of these states.
This makes energy conserved already
at the level of (\ref{e2}), making the analysis simpler. 
The product over $\omega$ shows that states of the form
$|n_{\omega}\rangle |n_{\omega'}\rangle \cdots$ with total
energies $E=\omega n_{\omega} + \omega' n_{\omega'} +\ldots$ also appear.
Thus, it is convenient to rewrite (\ref{e2}) as a sum over energy eigenstates
$|\pm E,\xi\rangle$
\begin{equation}\label{e3}
|\psi\rangle_{\rm squeeze}=\sum_E\sum_{\xi}d_{E,\xi}(M)
|-E,\xi\rangle \otimes |E,\xi\rangle ,
\end{equation}
where $\xi$ labels different states having the same outside or 
inside energy $\pm E$, and the sum is taken over non-negative values
of $E$.
The coefficients $d_{E,\xi}$ can be expressed in terms of $f_{\omega,n}$,
but the explicit expression will not be needed here.
The squeezed state (\ref{e2}) is a pure state and the transition
(\ref{e1}) is unitary \cite{gris}. Consequently, the density matrix
constructed from (\ref{e3}) is pure.
However, an outside observer cannot observe the 
inside states, so the density matrix describing the knowledge of the
outside observer is given by tracing out the inside degrees of freedom
of the total density matrix. Applying this to
(\ref{e3}), one obtains 
\begin{equation}\label{out}
\rho_{\rm out}=\sum_E\sum_{\xi}|d_{E,\xi}(M)|^2 \,
|E,\xi\rangle \langle E,\xi| ,
\end{equation}
which is a mixed state. 
However, we have argued that the negative-energy states are not physical,
which means that the mixed thermal state (\ref{out}) 
is obtained by tracing out over unphysical degrees of freedom. Hence,
this mixed thermal state may also be unphysical. A physical density matrix
should be obtained by tracing out over physical (but unobserved) 
degrees of freedom. The difference between unphysical and unobserved
degrees is in the fact that the former cannot be observed even in principle,
by any observer.

The unphysical negative-energy particles can be intuitively viewed
as virtual particles analogous to those appearing in Feynman diagrams
of conventional perturbative QFT. They cannot exist as final measurable
states. Instead, they must be {\em absorbed} by physical states. In our case,
the physical object that should absorb them is the black hole.
To give a precise description of this process of absorption, one should
invoke a precise microscopic theory that presumably includes a
quantum theory of gravity as well. Nevertheless, the essential features of such
an absorption can be understood even without a precise microscopic theory.
For simplicity, we study uncharged and unrotating black holes. Thus, 
we assume that a black hole with a mass $M$ can be described by a
quantum state $|M;\chi_M\!>$, where $\chi_M$ labels different BH states
having the same mass $M$. We assume that the number of different states
increases with $M$ and that there is only one state with mass $M=0$,
i.e., that $|0;\chi_0\!>=|0\!>$. In particular, such an assumption
is consistent with string theory asserting that entropy 
of the internal BH degrees of freedom is proportional to the surface,
i.e., to $M^2$. It is also consistent with a more naive possibility
that the entropy is proportional to the volume, i.e., to $M^3$.
In fact, proportionality of entropy to the surface rests
on the validity on the Einstein equation, while thermal particle creation
from a horizon
is a much more general phenomenon \cite{viss}. As our analysis will not 
depend on validity of the Einstein equation, we will not be able to specify
the exact number of states with mass $M$. For our purposes, it is sufficient
to assume that the absorption of negative-energy particles takes a generic form
\begin{equation}\label{e4}
|M;\chi_M\!> |-E,\xi\rangle \stackrel{\rm absorp}{\longrightarrow}
|M-E;\chi_{M-E}\!> .
\end{equation}
Such a form is dictated by energy conservation, which, indeed, is consistent
with the first law of BH thermodynamics.
Note that the left-hand side of (\ref{e4}) has a larger number 
of different states than the right-hand side. Consequently, the operator 
governing the absorption (\ref{e4}) is not invertible, and thus cannot be
unitary. Nevertheless, the overall unitarity is not necessarily violated.
To see why, note that, although the squeezing (\ref{e1}) is  
described by a formally unitary operator, it is not unitary on the
{\em physical} Hilbert space (because the physical Hilbert space
does not contain the unphysical negative-energy particles). 
Thus, neither the squeezing (\ref{e1}) nor the absorption (\ref{e4}) are
physical processes by themselves. What is physical is their composition
\begin{equation}\label{e5}
|M;\chi_M\!> \rightarrow \sum_E\sum_{\xi}d_{E,\xi}(M)
|M-E;\chi_{M-E}\!> \otimes |E,\xi\rangle . 
\end{equation}
Thus, if the initial state is $|\Psi_0\rangle=|M;\chi_M\!>$, 
then we have a physical transition
$|\Psi_0\rangle \rightarrow |\Psi_1\rangle$, where 
$|\Psi_1\rangle$ is the right-hand side of (\ref{e5}).
The physical process (\ref{e5}) is expected to be unitary.
(An explicit verification
of unitarity requires a more specific model of quantum gravity.)
In fact, one may forget about the virtual subprocesses (\ref{e1}) and (\ref{e4})
and consider (\ref{e5}) as the only directly relevant physical process.
Indeed, the process of BH radiation in a more advanced theory of quantum gravity may not be based on a Bogoliubov transformation at all, so it may not
be formulated in terms of creation of virtual negative-energy particles
appearing in (\ref{e1}), but directly in terms of physical processes
of the form of (\ref{e5}). In fact, this is exactly what occurs
in string theory \cite{das}.
 
Note also that in (\ref{e4}) we assume that the right-hand side
does not depend on $\xi$. This reflects on the right-hand side
of (\ref{e5}) in the fact that the new BH state does not depend on the
state of radiation $\xi$. This means that there is no correlation between
radiated particles and BH interior, except for the trivial correlation
expressing the fact that total energy must be conserved. 
The absence of such correlations is expected also from a 
more general view of the semiclassical description
of particle creation \cite{niksc}.
As we shall see,
this destruction of the (unphysical) information contained in the
negative-energy particles on the left-hand side of (\ref{e4})
makes the remnant scenario viable, by removing the unwanted huge information
that otherwise would have to be be present in a light remnant.
Nevertheless, later we also discuss a possibility to relax the 
assumption that the nontrivial correlation between exterior
radiation and BH interior is completely absent.

\section{The process of radiation -- unitary evolution
and the role of wave-function collapse}

Now the analysis of further steps of the process of BH radiation is
mainly technical. After (\ref{e5}), the remaining BH state radiates
again, now at a new larger temperature corresponding to the new smaller
BH mass $M-E$. Thus, the next step $|\Psi_1\rangle \rightarrow |\Psi_2\rangle$
is based on a process analogous to (\ref{e5}) 
\begin{eqnarray}\label{e6}
|M-E;\chi_{M-E}\!> \rightarrow \sum_{E'}\sum_{\xi'}d_{E',\xi'}(M-E)
\nonumber \\
 \times |M-E-E';\chi_{M-E-E'}\!> \otimes |E',\xi'\rangle , 
\end{eqnarray} 
so
\begin{eqnarray}\label{e7}
|\Psi_2\rangle  = \sum_E\sum_{E'}\sum_{\xi}\sum_{\xi'} 
d_{E,\xi}(M) d_{E',\xi'}(M-E) \nonumber \\
 \times |M-E-E';\chi_{M-E-E'}\!> \otimes |E,\xi\rangle |E',\xi'\rangle . 
\end{eqnarray}
Repeating the same process $t$ times, we obtain
\begin{eqnarray}\label{e8}
|\Psi_t\rangle & = & \sum_{E_1}\cdots\sum_{E_t}
\sum_{\xi_1}\cdots\sum_{\xi_t} \nonumber \\
& \times &
d_{E_1,\xi_1}(M) \cdots d_{E_t,\xi_t}(M-E_1- \cdots -E_{t-1}) \nonumber \\
& \times & 
|M-{\cal E};\chi_{M-{\cal E}}\!> \otimes |E_1,\xi_1\rangle \cdots |E_t,\xi_t\rangle ,
\end{eqnarray}
where ${\cal E}=\sum_{t'=1}^{t}E_{t'}$.
(A continuous description of evolution labeled by a continuous time parameter $t$ is also possible, but this does not change our main conclusions.)
States with the same energy ${\cal E}$ can be grouped together,
so we can write 
\begin{equation}\label{e10}
|\Psi_t\rangle=\sum_{\cal E} \sum_{\Xi} |M-{\cal E};\chi_{M-{\cal E}}\!> \otimes 
D^{(t)}_{{\cal E},\Xi} |{\cal E},\Xi\rangle ,
\end{equation}
where $\Xi=\{ \xi_1,\ldots, \xi_t \}$ and the coefficients
$D^{(t)}_{{\cal E},\Xi}$ can be expressed in terms of
$d_{E_{t'},\xi_{t'}}$.
Note that,
for any finite $t$, $|\Psi_t\rangle$ contains contributions
from all possible BH masses $M'=M-{\cal E}$. At first sight,
it seems to imply that the unitary evolution (\ref{e10}) prevents the
black hole from evaporating completely during a finite time $t$.
Nevertheless, this is not really true. To see why, it is instructive
to consider a simpler quantum decay $|a\rangle\rightarrow|b\rangle$ 
in which the unitary evolution usually implies an exponential law
$|\psi(t)\rangle=\sqrt{1-e^{-\Gamma t}}|b\rangle + \sqrt{e^{-\Gamma t}}|a\rangle$.
For any finite $t$, there is a finite probability $e^{-\Gamma t}$
that the decay has not yet occurred. Nevertheless, a wave-function 
collapse associated with an appropriate quantum measurement implies
that at each time the particle will be found either in the state
$|a\rangle$ or $|b\rangle$. Analogously, if the BH mass
$M'$ is measured at time $t$, the wave-function collapse implies
\begin{equation}\label{e11}
|\Psi_t\rangle \stackrel{\rm measure}{\longrightarrow}
|M-{\cal E};\chi_{M-{\cal E}}\!> \otimes
N_{\cal E} \sum_{\Xi} D^{(t)}_{{\cal E},\Xi} |{\cal E},\Xi\rangle ,
\end{equation}
where $N_{\cal E}$ is the normalization factor, 
$N_{\cal E}^{-2}=\sum_{\Xi} |D^{(t)}_{{\cal E},\Xi}|^2$.
Now the black hole is in a definite pure state
$|M-{\cal E};\chi_{M-{\cal E}}\!>$ and the outside particles are in a
definite pure state
$N_{\cal E} \sum_{\Xi} D^{(t)}_{{\cal E},\Xi} |{\cal E},\Xi\rangle$.
(More realistically, the measurement uncertainty $\Delta M'$ is smaller
for smaller $M'$, so the outside particles are closer to a pure state
when $M'$ is smaller.) For example,
it is conceivable that some quantum mechanism might prevent transitions
(\ref{e5}) for $M-E<M_{\rm min}$ (where $M_{\rm min}$ is a hypothetic
minimal possible BH mass). 
In this case, (\ref{e11}) may correspond
to a transition to a BH remnant with a mass $M-{\cal E}=M_{\rm min}$.
Such a BH remnant is not correlated with the radiated particles
(except for the correlation implied by energy conservation)
and the information content of the remnant is determined only by its mass.
The absence of such correlations is a consequence of the assumption
that the right hand-side of (\ref{e4}) does not depend on $\xi$.
This assumption could also be relaxed by allowing that at least
some different $\xi$'s may correspond to different BH states.
In this case, the BH state in (\ref{e11}) would also depend
on $\Xi$, so it would not sit in front of the sum over $\Xi$,
which would imply that neither the black hole nor the radiation
is in an exactly pure state, but that there is a small correlation
between them. Nevertheless, the maximal amount of possible correlation is restricted by the smallness of the BH mass. In particular, if 
$M_{\rm min}=0$, then (\ref{e11}) may correspond to a complete
evaporation of the black hole, in which case the BH state
$|0\!>$ must be unique, implying that the final state of radiation
must be a pure state
$N_M \sum_{\Xi} D^{(t)}_{M,\Xi} |M,\Xi\rangle$.

\section{Discussion -- thermality, apparent nonunitarity, and 
the origin of nonlocality}

We have seen that, under reasonable assumptions, the BH radiation is in a 
pure state whenever the BH mass is measured exactly. Does it mean that
the BH radiation is not really thermal? Actually not. 
Instead, the situation is analogous
to that in the discussion around Eqs.~(\ref{e2.1})-(\ref{decoh}).
For example, if the BH mass is measured after the first step
(\ref{e5}), then the radiation collapses to a pure state
equal (up to an overall normalization factor) to 
$\sum_{\xi} d_{E,\xi}(M) |E,\xi\rangle$. This state is obtained from 
$\prod_{\omega}\sum_n f_{\omega,n}(M)|n_{\omega}\rangle$
by rewriting it as a sum of products and retaining only those states
the total energy of which is equal to $E$. The density matrix
of such a pure state takes a form analogous to (\ref{rho}). 
Due to the decoherence induced by the interaction with the environment,
in practice such a state can be effectively described by a mixed state
analogous to (\ref{decoh}). From (\ref{e2.1}) we see that it is a thermal
mixed state. More precisely, as the total energy $E$ is exactly 
specified, while the number of particles is specified only in average, 
this is a thermal state corresponding to a grand microcanonical ensemble.
By contrast, the thermal state (\ref{out}) (in which both total energy and number
of particles are specified only in average) corresponds to a
grand canonical ensemble.

At the end, let us recall that our resolution of the BH information
puzzle involves 4 different types of seemingly nonunitary evolutions.
The process of squeezing (\ref{e1}) is formally unitary \cite{gris},
but it is not unitary on the physical space. It is allways accompanied
with another nonunitary virtual process (the absorption
of negative-energy particles) Eq.~(\ref{e4}), which together are combined
into a physical unitary process (\ref{e5}). This represents the core of
our resolution of the BH information puzzle. The third nonunitary process
is the wave-function collapse (\ref{e11}). The exact meaning of 
the collapse depends on the general interpretation of QM that one adopts.
In particular, in some interpretations (e.g., many-world interpretation
and the Bohmian interpretation) a true collapse does not really exist,
making QM fully consistent with unitarity. Finally, the fourth
nonunitary process is the phenomenon of decoherence (\ref{decoh}), 
which corresponds only to an effective violation of unitarity, not a fundamental
one. 

Finally note that, although our analysis allows a complete BH evaporation
without a true violation of unitarity, no new nonlocal mechanism
has been involved. The only new mechanism is the absorption
(\ref{e4}), which, however, occurs only inside the black hole,
thus not violating locality. Some nonlocal mechanisms {\em are} involved
in our analysis, namely quantum entanglement and quantum 
wave-function collapse, but these are standard nonlocal aspects of QM.  

\section*{Acknowledgements}

The author is grateful to T.~Jacobson for valuable discussions
on Hawking radiation and BH information.
This work was supported by the Ministry of Science of the
Republic of Croatia under Contract No.~098-0982930-2864.

\end{document}